\documentclass[conference,a4paper,10pt]{IEEEtran}

\usepackage{amsmath}
\usepackage{amssymb}

\makeatletter
\let\MYcaption\@makecaption
\makeatother

\usepackage[font=footnotesize]{subcaption}

\makeatletter
\let\@makecaption\MYcaption
\makeatother

\usepackage{pgfplots}
\pgfplotsset{compat=1.15}

\definecolor{color0}{rgb}{0.12156862745098,0.466666666666667,0.705882352941177}
\definecolor{color1}{rgb}{1,0.498039215686275,0.0549019607843137}
\definecolor{color2}{rgb}{0.172549019607843,0.627450980392157,0.172549019607843}
\definecolor{color3}{rgb}{0.83921568627451,0.152941176470588,0.156862745098039}

\definecolor{match}{rgb}{.565, .933, .565} 		
\definecolor{overest}{rgb}{1., .60, .60}			
\definecolor{underest}{rgb}{.678, .847, .902} 	

\usepackage{todonotes}

\usepackage{siunitx}
\usepackage{booktabs}
\usepackage{colortbl}

\DeclareSIUnit{\sample}{S}

\DeclareMathOperator{\besseli}{I}

\usepackage[style=ieee,url=false,isbn=false,maxbibnames=6]{biblatex}
\AtBeginBibliography{\footnotesize}

\usepackage{ifthen}

\makeatletter
\newcounter{IEEE@bibentries}
\renewcommand\IEEEtriggeratref[1]{%
  \renewbibmacro{finentry}{%
    \stepcounter{IEEE@bibentries}%
    \ifthenelse{\equal{\value{IEEE@bibentries}}{#1}}
    {\finentry\@IEEEtriggercmd}
    {\finentry}%
  }%
}
\makeatother

\usepackage[abbreviations,symbols,automake]{glossaries-extra}
\setabbreviationstyle[acronym]{long-short}
\makeglossaries
\newacronym{3gpp}{3GPP}{3rd Generation Partnership Project}
\newacronym{5g}{5G}{fifth generation mobile networks}
\newacronym{6g}{6G}{sixth generation mobile networks}
\newacronym{bs}{BS}{base station}
\newacronym{cir}{CIR}{channel impulse response}
\newacronym{cdf}{CDF}{cumulative distribution function}
\newacronym{ctf}{CTF}{channel transfer function}
\newacronym{dft}{DFT}{discrete Fourier transform}
\newacronym{dl}{DL}{downlink}
\newacronym{dkw}{DKW}{Dvoretzky-Kiefer-Wolfowitz}
\newacronym{fec}{FEC}{forward error correction}
\newacronym{idft}{IDFT}{inverse discrete Fourier transform}
\newacronym{iid}{iid}{independent identically distributed}
\newacronym{inid}{inid}{independent non-identically distributed}
\newacronym{los}{LOS}{line of sight}
\newacronym{lpda}{LPDA}{Log-Periodic Dipole Array}
\newacronym{lsas}{LSAS}{large scale antenna system}
\newacronym{lte}{LTE}{Long Term Evolution}
\newacronym{mimo}{MIMO}{multiple input multiple output}
\newacronym{miso}{MISO}{multiple input single output}
\newacronym{mpc}{MPC}{multi path component}
\newacronym{mrc}{MRC}{maximal ratio combining}
\newacronym{ni}{NI}{National Instruments}
\newacronym{nlos}{NLOS}{non-line of sight}
\newacronym{ntnu}{NTNU}{Norwegian University of Technology and Science}
\newacronym{ofdm}{OFDM}{orthogonal frequency division multiplexing}
\newacronym{pdf}{PDF}{probability density function}
\newacronym{reranp}{ReRaNP}{Reconfigurable Radio Network Platform}
\newacronym{rms}{rms}{root mean square}
\newacronym{rx}{RX}{receiver}
\newacronym{sc}{SC}{selection combining}
\newacronym{simo}{SIMO}{single input multiple output}
\newacronym{siso}{SISO}{single input single output}
\newacronym{snr}{SNR}{signal to noise ratio}
\newacronym{tdd}{TDD}{time division duplex}
\newacronym{tx}{TX}{transmitter}
\newacronym{twdp}{TWDP}{two-wave diffuse power}
\newacronym{ue}{UE}{user equipment}
\newacronym{ul}{UL}{uplink}
\newacronym{wsn}{WSN}{wireless sensor network}
\newacronym{ur}{UR}{ultra-reliability}
\newacronym{urllc}{URLLC}{ultra-reliable low-latency communication}
\newacronym{ecdf}{ECDF}{empirical cumulative distribution function}

\newcommand{\vect}[1]{\mathbf{#1}}
\newcommand{\matr}[1]{\mathbf{#1}}

\newcommand{\abs}[1]{\left|{#1}\right|}
\newcommand{\norm}[1]{\left\lVert{#1}\right\rVert}

\newcommand{\herm}[1]{{#1^\text{H}}}
\newcommand{\trans}[1]{{#1^\text{T}}}

\newcommand{\expect}[1]{\mathop{\mathbb{E}}\!\left\{{#1}\right\}}

\title{Local Diversity and Ultra-Reliable Antenna Arrays}
                                    
\author{
Jens~Abraham and %
Torbj{\"o}rn~Ekman %
\thanks{J. Abraham and T. Ekman are with the Department of Electronic Systems, Norwegian University of Science and Technology, Norway. 
e-mail: \texttt{\{jens.abraham, torbjorn.ekman\}@ntnu.no}}
}

\IEEEoverridecommandlockouts 

\begin{document}

\maketitle 
\widowpenalty10000 

\begin{abstract}
\Glsxtrlong{urllc} enables new use cases for mobile radio networks.
The \glsxtrfull{ur} regime covers outage probabilities between \num{e-9} and \num{e-5}, obtained under stringent latency requirements.
Characterisation of the \glsxtrshort{ur}-relevant statistics is difficult due to the rare nature of outage events, but diversity defines the asymptotic behaviour of the small-scale fading distributions' lower tail.
The \glsxtrshort{ur}-relevant regime in large-scale antenna systems behaves differently from the tail.
We present the generalising \emph{local diversity} at a certain outage probability to show this difference clearly. 
For more than four independent antenna elements, the classic diversity overestimates and underestimates the slope of the cumulative density function for weak and strong deterministic channel components, respectively.
\end{abstract}
\begin{IEEEkeywords}
channel hardening, massive MIMO, Rician fading, URLLC.
\end{IEEEkeywords}


\section{Introduction}

One of the reoccurring promises in both \gls{5g} and \gls{6g} specifications is \gls{urllc}.
The \gls{urllc} requires an outage probability of $10^{-5}$ or better within a \SI{1}{\milli\second} transmission period in \gls{5g} \cite{ji_ultra-reliable_2018}.
The authors of \cite{eggers_wireless_2019} introduce the terminology of \emph{\gls{ur}-relevant statistics} for outage probabilities below \num{e-5}.
It can be expected that the requirements for \gls{6g} will be even more stringent.
Hence, we will consider outage probabilities between \num{e-9} and \num{e-5} as the \emph{\gls{ur}-relevant regime}.

Generally, the allowed latency can be used to retransmit a packet, if the original message did not reach its destination.
By decreasing the permitted latency, only one-shot transmissions can ultimately fulfil the requirement because a retransmission would take too long.
This type of requirement is typical in control-loop or event based applications, where the timing is critical.
Alternatively, the age of information \cite{kosta_age_2017} can used as a design metric, where the state of a system is observed.
Here, a non-successful transmission transmission every now and then might be acceptable, since the system can cope with intermittent link failure.

Small-scale fading is one of the main reasons for link-loss in rich scattering environments.
It can be counteracted with \gls{fec}, relying on the assumption that fading events are short enough with respect to the coded packet length.
If the coherence time of the channel is longer than the latency requirement, alternative measures have to be used to overcome small-scale fading.
Exploiting spatial diversity through massive \gls{mimo} can improve the link robustness due to channel hardening. 
This approach reduces the variation of the channel gain around its mean and hereby the outage probability.
Recenctly, we have suggested to use a fading margin to characterise channel hardening \cite{abraham_fading_2021}. 
It describes the required excess gain to provide a certain outage probability at a chosen rate.
Hence, the performance of an \gls{ur} antenna array with varying number of antenna elements can be quantified clearly.

An additional caveat for \gls{urllc} is power limitation of users.
Especially battery powered sensors in \gls{wsn} should avoid retransmissions.
In those cases, minimising the fading margin improves the energy efficiency and allows to meet \gls{ur} target outage probabilities.
Moreover, smaller fading margins reduce the interference levels for users of the same system and systems that share the same spectrum resource.

This outlines why large antenna arrays are a technically viable solution for narrow-band \gls{urllc} without retransmission of packets.
System level simulations based on a \gls{3gpp} channel model for a specific cell show promising results for a coherence interval based pilot strategy \cite{yan_can_2021}.

A fundamental question remains, how can we infer the system behaviour of events that barely ever happen?
A neat approach is the characterisation of the lower tail of the \gls{cdf} as an intermediate solution between parametric channel models and non-parametric models \cite{angjelichinoski_statistical_2019}.
The lower tail of multiple common fading distributions follows a power law \cite{eggers_wireless_2019}, which gives the possibility to relax the model assumption from a single distribution to a class of distributions.
The power law approximation requires two parameters: an offset and the log-log slope of the \gls{cdf}.
E.g. the classic Rayleigh channel shows a well known slope of \SI{10}{\deci\bel} per decade in the lower tail.

Furthermore, the outage probability in detection problems \cite{tse_fundamentals_2005} for high \gls{snr} corresponds to the lower tail of the channel gain.
Using the \gls{snr} emphasises the variation introduced due to the small-scale fading channel and avoids a dependency on a specific modulator and detector.
Due to that correspondence, the log-log slope in the asymptotic lower tail reveals the diversity of the radio channel.
We propose to evaluate the log-log slope at a specific probability, generalising it to the \emph{local diversity}.
Hence, for outage probabilities converging to zero it attains the classic diversity measure. 

A dual slope behaviour in single antenna Rician fading channels with larger $\mathcal{K}$-factors has already been shown in \cite{hon_tat_hui_performance_2005}.
For multi-antenna systems in both Rayleigh and Rician fading, the outage probability slope in the \gls{ur}-relevant regime deviates from the classic diversity.
Therefore, a power law approximation of the lower tail can not provide an accurate description of the \gls{cdf} for massive \gls{mimo} systems.

\emph{Our main contribution is the local diversity to highlight that lower tail approximations do not cover the actual system behaviour in the \gls{ur}-relevant regime.}
We motivate the usage of analytical tools to get insight into \gls{ur}-relevant statistics in the next section, because the number of of necessary observations for a reliable empirical approach is prohibitive for real world scenarios.
An uncorrelated Rician multi-antenna fading environment is introduced in section \ref{sec:rice_fading}.
It's local diversity is derived to relate the classical diversity to the \gls{ur}-relevant regime.
This measure can be seen as the relative error of a power law approximation based on the asymptotic behaviour of the lower tail.
New compact expressions for the \gls{cdf}, \gls{pdf} and local diversity in terms of the complementary Marcum-Q function are used to evaluate them for large scale antenna systems.
We provide a comparison of multi-antenna systems in different Rician fading environments with respect to the fading margin in section \ref{sec:discussion}, to discuss the scaling behaviour.
Furthermore, sampling strategies to analyse the \gls{ur}-relevant regime are outlined.

\section{Predicting the Unpredictable?} \label{sec:ecdf_observation}

Let us investigate \glspl{ecdf} as non-parameteric model, to understand the value of parametric analytical models for \gls{ur}-relevant statistics.
Basically, the \gls{ur}-relevant regime covers the behaviour of rare events that barely ever happen and the fewer assumptions necessary the more general is the solution.
How many observations are necessary to reliably estimate the \gls{ur}-relevant statistics without prior knowledge?

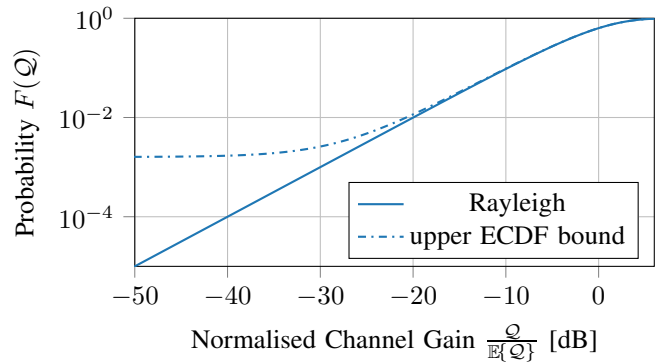
\begin{figure}
	\begin{tikzpicture}
\begin{axis}[
legend pos=south east,
tick align=outside,
tick pos=left,
xlabel={Normalised Channel Gain $\frac{\mathcal{Q}}{\expect{\mathcal{Q}}}$ [dB]},
xmajorgrids,
xmin=-50, xmax=6,
ylabel={Probability $F(\mathcal{Q})$},
ymode=log,
ymajorgrids,
ymin=1e-5, ymax=1,
width=.95\linewidth, height=.55\linewidth,
]
\addplot [thick, color0] table[x index=0, y index=2] {data/K_-1000-M_1.csv};
\addlegendentry{Rayleigh};
\addplot [thick, color0, dash dot] table[x index=0, y expr=\thisrowno{2}+1.6e-3] {data/K_-1000-M_1.csv};
\addlegendentry{upper \acrshort{ecdf} bound};
\end{axis}
\end{tikzpicture}
	\caption{The \gls{cdf} of a Rayleigh fading channel and an upper bound for an \gls{ecdf} is shown. The error term in Eqn. \eqref{eqn:dkw_bound} for a million observations and a confidence interval of \SI{99}{\percent} is used as example, showing that the estimation of outage probabilities below \num{1.6e-3} is unreliable.}
	\label{fig:dkw_cdf}
\end{figure}

The \gls{dkw} inequality \cite{dvoretzky_asymptotic_1956,massart_tight_1990} can be used to bound an $R$-sample \gls{ecdf} with respect to the true underlying \gls{cdf} leading to the error term $\epsilon$ with confidence $\xi$:
\begin{equation}
    \epsilon = \sqrt{\frac{\ln{\frac{2}{1-\xi}}}{2R}}. \label{eqn:dkw_bound}
\end{equation}
This error term is characterising an error floor for the \gls{ecdf} at low probabilities.
Taking $R = \num{e6}$ observations as example and aiming at a confidence of $\xi = \SI{99}{\percent}$ gives an error term of \num{1.6e-3}.
The resulting upper bound of the \gls{ecdf} for a true single-antenna Rayleigh fading channel is shown in Fig. \ref{fig:dkw_cdf}.
It can be seen that the \gls{ecdf} in the \gls{ur}-relevant regime would be much smaller than the error floor, rendering empirical estimation of outage probabilities below \num{1.6e-3} practically useless.

The number of antenna elements in massive \gls{mimo} ranges from a few ten to a few hundred, that can provide potentially correlated parallel observations of the radio channel.
The remaining observations have to be gathered in a stationary time-frequency window to belong to the same underlying \gls{cdf}.
This is very unlikely in realistic scenarios, especially for high (environmental) mobility with limited temporal stationarity.
Eventually, the characterisation of \gls{ur}-relevant statistics in the lower tail is prone to large estimation errors for non-parametric models.
Additionally, if energy efficient users are required, less spectrum may be used, reducing the number of samples in the spectral domain.
Hence, the spatial domain sampling provided by an antenna array has to provide both the robustness of the system as well as a number of observations to estimate the \gls{cdf}.

The large number of observations an obstacle even for simulations.
Assuming that outage probabilities of \num{e-6} with confidence of \SI{99.9999}{\percent} are of interest, on the order of \num{e13} observations have to be collected.
Both, runtime and memory requirements of Monte Carlo simulations become cumbersome to get reliable results for the \gls{ecdf}.
Hence, only the analytic study of the \gls{ur}-relevant regime has the possibility to give insight into trade-offs, as long as the model assumptions are not violated.


\section{Rician Fading Channel Revisited} \label{sec:rice_fading}

We will consider Rician fading channels with a Rician $\mathcal{K}$-factor and a diffuse power (gain) $P_\text{dif}$, following the parametrisation in \cite{durgin_space-time_2003}.
The $\mathcal{K}$-factor describes the ratio between a deterministic component and the diffuse power of the radio channel.
Hence, the mean power gain is $(\mathcal{K}+1) P_\text{dif}$. 

To take $M$ uncorrelated antennas at the base station into account, a complex random vector with mean $\sqrt{\mathcal{K} P_\text{dif}} \left[ e^{j \varphi_1}, e^{j \varphi_2}, \cdots, e^{j \varphi_M} \right]^T$ and covariance $P_\text{dif} \matr{I}$ is constructed:
\begin{equation}
	\vect{h} \in \mathbb{C}^M \sim \mathcal{CN} \left( \sqrt{\mathcal{K} P_\text{dif}} \left[ e^{j \varphi_1}, e^{j \varphi_2}, \cdots, e^{j \varphi_M} \right]^T , P_\text{dif} \matr{I} \right). \label{eqn:cn_h}
\end{equation}
The phases $\varphi_m$ represent the phase front of the deterministic component with respect to the antennas and $\matr{I}$ is the $M \times M$ identity matrix.
For Rayleigh fading ($\mathcal{K} = 0$), $\vect{h}$ is a circular-symmetric complex normal random vector $\vect{h} \sim \mathcal{CN} \left(0, P_\text{dif} \matr{I} \right)$.

The effective channel $\mathcal{H}$ for a \gls{mrc} weight vector $\vect{w}$ at the receiver results in:
\begin{equation}
	\mathcal{H} = \trans{\vect{w}} \vect{h} = \frac{\herm{\vect{h}}\vect{h}}{\sqrt{\norm{\vect{h}}_2^2}}= \sqrt{\sum_{m=1}^M \left| h_m \right|^2} \label{eqn:effective_channel_gain}.
\end{equation}

The \gls{cdf} $F(\mathcal{Q}; P_\text{dif}, \mathcal{K}, M)$ of the effective power gain $\mathcal{Q} = \abs{\mathcal{H}}^2$ of this multi-antenna Rician channel is compactly given by:
\begin{equation}
	F(\mathcal{Q}; P_\text{dif}, \mathcal{K}, M) = P_M \left( \mathcal{K} M, \frac{\mathcal{Q}}{P_\text{dif}} \right), \label{eqn:cdf}
\end{equation}
where $P_M(\cdot)$ is the complementary Marcum Q-function \cite{gil_algorithm_2014} with definition\footnote{Note that this definition is a different variant of the implementation found in major numeric computing environments, but the reference \cite{gil_algorithm_2014} provides a Fortran implementation together with the numerical algorithm description.}:
\begin{equation}
    P_\mu(x,y) = x^{\frac{1}{2} (1-\mu)} \int_0^{y} t^{\frac{1}{2}(\mu - 1)} e^{- t - x} \besseli_{\mu-1} \left( 2 \sqrt{x t} \right) dt.\label{eqn:marcum_p}
\end{equation}
This power gain \gls{cdf} is a generalised \cite{stacy_generalization_1962} or non-central gamma distribution \cite{ruben_non-central_1974} arising from a sum over squared per-antenna channel coefficients in Eqn. \eqref{eqn:effective_channel_gain}.

\begin{figure}
\centering
\begin{subfigure}[b]{0.99\linewidth}
\centering
\begin{tikzpicture}
\begin{axis}[
legend pos=south east,
tick align=outside,
tick pos=left,
xlabel={Normalised Channel Gain $\frac{\mathcal{Q}}{\expect{\mathcal{Q}}}$ [dB]},
xmajorgrids,
xmin=-30, xmax=6,
ylabel={Probability $F(\mathcal{Q})$},
ymode=log,
ymajorgrids,
ymin=1e-6, ymax=1,
width=.95\linewidth, height=.55\linewidth,
]
\addplot [thick, color0] table[x index=0, y index=2] {data/K_-1000-M_1.csv};
\addlegendentry{$-\infty\,\si{\deci\bel}$};
\addplot [thick, color1] table[x index=0, y index=2] {data/K_0-M_1.csv};
\addlegendentry{$\SI{0}{\deci\bel}$};
\addplot [thick, color2] table[x index=0, y index=2] {data/K_3-M_1.csv};
\addlegendentry{$\SI{3}{\deci\bel}$};
\addplot [thick, color3] table[x index=0, y index=2] {data/K_10-M_1.csv};
\addlegendentry{$\SI{10}{\deci\bel}$};
\end{axis}
\end{tikzpicture}
\caption{\glspl{cdf}}
\label{fig:cdfs}
\end{subfigure}

\begin{subfigure}[b]{0.99\linewidth}
\centering
\begin{tikzpicture}
\begin{axis}[
legend pos=north west,
tick align=outside,
tick pos=left,
xlabel={Normalised Channel Gain $\frac{\mathcal{Q}}{\expect{\mathcal{Q}}}$  [dB]},
xmajorgrids,
xmin=-30, xmax=6,
ylabel={Local Diversity $\mathcal{D}(\mathcal{Q})$},
ymajorgrids,
ymin=0, ymax=4,
width=.95\columnwidth, height=.55\columnwidth,
]
\addplot [thick, color0] table[x index=0, y index=3] {data/K_-1000-M_1.csv};
\addlegendentry{$-\infty\,\si{\deci\bel}$};
\addplot [thick, color1] table[x index=0, y index=3] {data/K_0-M_1.csv};
\addlegendentry{$\SI{0}{\deci\bel}$};
\addplot [thick, color2] table[x index=0, y index=3] {data/K_3-M_1.csv};
\addlegendentry{$\SI{3}{\deci\bel}$};
\addplot [thick, color3] table[x index=0, y index=3] {data/K_10-M_1.csv};
\addlegendentry{$\SI{10}{\deci\bel}$};
\end{axis}
\end{tikzpicture}
\caption{local diversity}
\label{fig:loc_div}
\end{subfigure}
\caption{The normalised single antenna Rician channel is displayed for different $\mathcal{K}$-factors. The normalisation enforces unit mean. Larger $\mathcal{K}$-factors lead to a dual slope \gls{cdf}. The steeper slope corresponds to the superelevation of the local diversity.}
\end{figure}
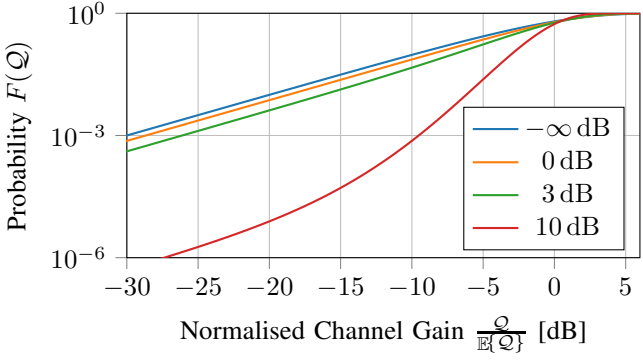
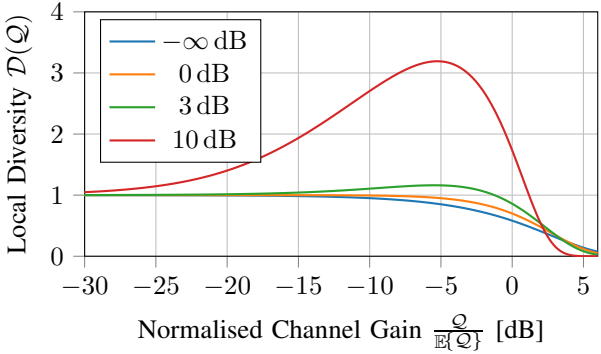

The distribution relates to a $\kappa$-$\mu$ envelope distribution \cite{yacoub_kappa_mu_2007}, where the number of independent antenna elements corresponds to $\mu$ clusters and the $\mathcal{K}$-factor relates to the ratio $\kappa$ between dominant and scattered channel components for a mean normalised to unity.
The connection between a single antenna Rayleigh channel, the complementary Marcum Q-function and the effective gain \gls{cdf} to arrive at a non-central gamma distribution is described in detail in the appendix \ref{sec:app-cdf} and the connection to the $\kappa$-$\mu$ envelope distribution follows directly from comparison of the \glspl{cdf}.

The mean effective power gain is:
\begin{equation}
	\expect{\mathcal{Q}} = M (\mathcal{K} + 1) P_\text{dif} \label{eqn:mean_power_gain},
\end{equation}
which follows from adding $M$ independent Rician channels with the same $\mathcal{K}$-factor and power in the diffuse component.
Varying $\mathcal{K}$-factors for different antenna elements could be accounted for, by using the mean $\mathcal{K}$-factor in the above formulation.
Both the $\mathcal{K}$-factor and the number of antenna elements $M$, have similar influence on the mean of the distribution.

Fig. \ref{fig:cdfs} shows a selection of \glspl{cdf} that describe the behaviour of a single antenna Rice channel.
The channel gain is normalised with its mean to allow easier comparison of the small-scale fading aspects for different $\mathcal{K}$-factors.
A stronger deterministic component leads to a dual slope behaviour with a steeper gradient closer to the median of the distribution.
Nonetheless, the gradient converges to \SI{10}{\deci\bel} per decade in the lower tail and is independent of the $\mathcal{K}$-factor.
The very seldom cases occur only when the diffuse components can cancel the deterministic component almost perfectly.
For a $\mathcal{K}$-factor of \SI{10}{\deci\bel}, the gradient is steepest in the region between \SI{-15}{\deci\bel} and \SI{0}{\deci\bel} with respect to the mean.
This indicates that the lower tail approximation underestimates the channel behaviour for outage probabilities ranging from \num{e-4} to \num{0.5}.

For sake of completeness is the corresponding \gls{pdf} $f(\mathcal{Q})$ of the effective power gain given in the following equation.
\begin{align}
	f&(\mathcal{Q}; P_\text{dif}, \mathcal{K}, M) \notag \\
	&= \begin{cases} \frac{1}{P_\text{dif}} e^{ - \frac{\mathcal{Q}}{P_\text{dif}} - \mathcal{K}}  \besseli_{0}\left( 2 \sqrt{ \mathcal{K} \frac{\mathcal{Q}}{P_\text{dif}}}\right) & M=1 \\
	\frac{1}{P_\text{dif}} \left( P_{M-1}(\mathcal{K}M, \frac{\mathcal{Q}}{P_\text{dif}}) - P_{M}(\mathcal{K}M, \frac{\mathcal{Q}}{P_\text{dif}})\right) & M > 1.
 \end{cases} \label{eqn:pdf}
\end{align}
Here, $I_0(\cdot)$ is the zero-order modified Bessel function of the first kind.
For the multi-antenna case, we can exploit the relation for derivatives of the complementary Marcum-Q function \cite[Sec. 2.3]{gil_algorithm_2014}.

\subsection{Local Diversity} \label{ssec:local_diversity}

So far, the local diversity has only been introduced conceptually.
Let us recall a common rule of thumb: the outage probability of a single antenna Rayleigh fading channel scales with \SI{10}{\deci\bel} per decade in the lower tail.
Furthermore, we have observed that a single antenna in narrowband Rician fading provides a diversity of one, too.

Therefore, a slope of \SI{10}{\deci\bel} per decade outage probability is used as reference and we define the local diversity as derivative of the scaled logarithmic \gls{cdf} of the channel power gain $\mathcal{Q}$ in \si{\deci\bel}:
\begin{equation}
    \mathcal{D}(\mathcal{Q}) = \frac{\partial}{\partial 10^{\mathcal{Q}/10}} 10 \log_{10} \left( F(\mathcal{Q}) \right) = \mathcal{Q} \frac{f(\mathcal{Q})}{F(\mathcal{Q})}. \label{eqn:ld_def}
\end{equation}
This ensures a scaling of $10/D$ \si{\deci\bel} per decade outage probability locally at $\mathcal{Q}$.
E.g. a local diversity of 10, 33 and 100 describes a slope of \SI{1}{\deci\bel}, \SI{.3}{\deci\bel} and \SI{.1}{\deci\bel} per decade outage probability, respectively.
The classic diversity is attained by evaluating the local diversity for $\mathcal{Q} \rightarrow -\infty\,\si{\deci\bel}$.

Resolving the differentiation in Eqn. \eqref{eqn:ld_def} reveals the quotient between \gls{pdf} $f(\mathcal{Q})$ and \gls{cdf} $F(\mathcal{Q})$, also known as inverse Mills' ratio, multiplied with $\mathcal{Q}$.
To study how well a lower tail approximation represents the behaviour of the radio channel in the \gls{ur}-relevant region for Rician channels, we use Eqns. \eqref{eqn:pdf} and \eqref{eqn:cdf} for the \gls{pdf} and \gls{cdf} of the effective power gain, respectively.
The local diversity for antenna arrays can be expressed in terms of the complementary Marcum-Q function for $M > 2$:
\begin{equation}
    \mathcal{D}(\mathcal{Q}; P_\text{dif}, \mathcal{K}, M) = \frac{\mathcal{Q}}{P_\text{dif}} \left(\frac{P_{M-1}(\mathcal{K}M, \frac{\mathcal{Q}}{P_\text{dif}})}{P_{M}(\mathcal{K}M, \frac{\mathcal{Q}}{P_\text{dif}})} - 1 \right).
\end{equation}

Fig. \ref{fig:loc_div} presents the local diversity for a single antenna Rician channel ($M=1$).
Larger $\mathcal{K}$-factors lead to a superelevated region before before convergence to unity.
The local diversity quantifies the increased steepness of the \glspl{cdf} in Fig. \ref{fig:cdfs}.
\begin{figure}[tb]
\centering
\begin{tikzpicture}
\begin{axis}[
legend pos=north west,
tick align=outside,
tick pos=left,
xlabel={Probability},
xmajorgrids,
xmin=1e-9, xmax=1,
xmode=log,
ylabel={Local Diversity $\mathcal{D}(\mathcal{Q})$},
ymajorgrids,
ymin=0, ymax=4,
width=.95\linewidth, height=.55\linewidth,
]
\addplot [thick, color0] table[x index=2, y index=3] {data/K_-1000-M_1.csv};
\addlegendentry{$-\infty\,\si{\deci\bel}$};
\addplot [thick, color1] table[x index=2, y index=3] {data/K_0-M_1.csv};
\addlegendentry{$\SI{0}{\deci\bel}$};
\addplot [thick, color2] table[x index=2, y index=3] {data/K_3-M_1.csv};
\addlegendentry{$\SI{3}{\deci\bel}$};
\addplot [thick, color3] table[x index=2, y index=3] {data/K_10-M_1.csv};
\addlegendentry{$\SI{10}{\deci\bel}$};
\end{axis}
\end{tikzpicture}
\caption{The local diversity with respect to the probability of a single antenna Rician channel for different $\mathcal{K}$-factors.}
\label{fig:prob_loc_div}
\end{figure}
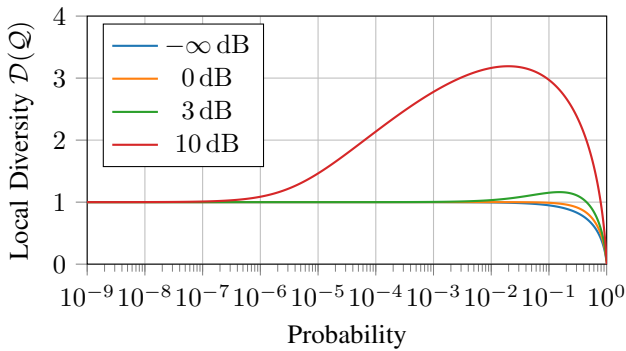
Fig. \ref{fig:prob_loc_div} plots the local diversity with respect to probability to interpret its behaviour in the \gls{ur}-relevant regime.
The superelevation is pronounced in the region from \num{e-6} to \num{0.5} for a $\mathcal{K}$-factor of \SI{10}{\deci\bel}.
All other $\mathcal{K}$-factors have converged to a local diversity of unity for probabilities smaller than \num{e-3}.

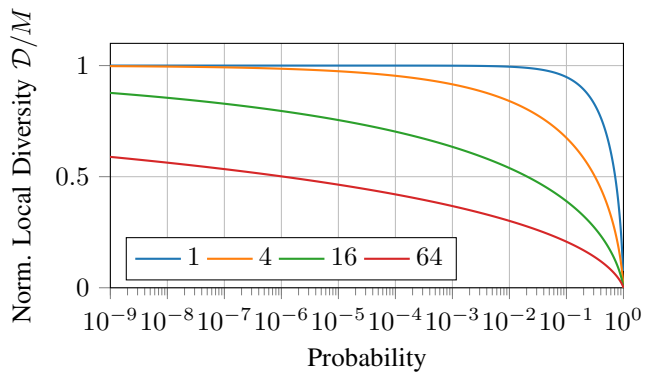
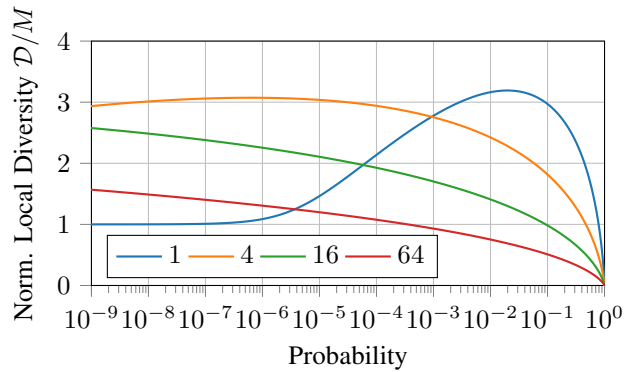
\begin{figure}
\centering
\begin{subfigure}[b]{0.99\linewidth}
\centering
\begin{tikzpicture}
\begin{axis}[
legend pos=south west,
legend columns=4, 
tick align=outside,
tick pos=left,
xlabel={Probability},
xmajorgrids,
xmin=1e-9, xmax=1,
xmode=log,
ylabel={Norm. Local Diversity $\mathcal{D}/M$},
ymajorgrids,
ymin=0, ymax=1.1,
width=.95\linewidth, height=.55\linewidth,
]
\addplot [thick, color0] table[x index=2, y expr=\thisrowno{3}/1] {data/K_-1000-M_1.csv};
\addlegendentry{$1$};
\addplot [thick, color1] table[x index=2, y expr=\thisrowno{3}/4] {data/K_-1000-M_4.csv};
\addlegendentry{$4$};
\addplot [thick, color2] table[x index=2, y expr=\thisrowno{3}/16] {data/K_-1000-M_16.csv};
\addlegendentry{$16$};
\addplot [thick, color3] table[x index=2, y expr=\thisrowno{3}/64]{data/K_-1000-M_64.csv};
\addlegendentry{$64$};
\end{axis}
\end{tikzpicture}
\caption{Rayleigh Fading}
\label{fig:loc_div_rayleigh}
\end{subfigure}

\begin{subfigure}[b]{0.99\linewidth}
\centering
\begin{tikzpicture}
\begin{axis}[
legend pos=south west,
legend columns=4, 
tick align=outside,
tick pos=left,
xlabel={Probability},
xmajorgrids,
xmin=1e-9, xmax=1,
xmode=log,
ylabel={Norm. Local Diversity $\mathcal{D}/M$},
ymajorgrids,
ymin=0, ymax=4,
width=.95\linewidth, height=.55\linewidth,
]
\addplot [thick, color0] table[x index=2, y expr=\thisrowno{3}/1] {data/K_10-M_1.csv};
\addlegendentry{$1$};
\addplot [thick, color1] table[x index=2, y expr=\thisrowno{3}/4] {data/K_10-M_4.csv};
\addlegendentry{$4$};
\addplot [thick, color2] table[x index=2, y expr=\thisrowno{3}/16] {data/K_10-M_16.csv};
\addlegendentry{$16$};
\addplot [thick, color3] table[x index=2, y expr=\thisrowno{3}/64]{data/K_10-M_64.csv};
\addlegendentry{$64$};
\end{axis}
\end{tikzpicture}
\caption{Rician Fading ($\mathcal{K} = \SI{10}{\deci\bel})$}
\label{fig:loc_div_rice}
\end{subfigure}
\caption{Normalised local diversity for an $M$-antenna array with 1, 4, 16 or 64 elements. A tail approximation would underestimate the outage behaviour of the system for larger arrays in Rayleigh fading and overestimate it in Rician fading.}
\end{figure}

\begin{table}[tb]
    \centering
    \caption{Normalised local diversity $\mathcal{D}/M$ evaluated at \num{e-6} probability. The different coloured regions show where the asymptotic tail approximation holds (green), underestimates (red) or overestimates (blue) the slope in the \gls{ur}-relevant regime.}
    \label{tab:nld_rice}
    \setlength{\tabcolsep}{4pt}
	\begin{tabular}{rSSSSSSSS}
	\toprule
	$\mathcal{K}$ & \multicolumn{7}{c}{Number of Antennas ($M$)} \\
	\cmidrule(lr){2-9}
	[\si{\deci\bel}] & {\centering 1} & {\centering 2} & {\centering 4} & {\centering 8} & {\centering 16} & {\centering 32} & {\centering 64} & {\centering 128} \\
	\midrule
	$-\infty$ &\cellcolor{match}  1.00 &\cellcolor{match}  1.00 &\cellcolor{match}  0.99 &\cellcolor{underest}  0.92 &\cellcolor{underest}  0.80 &\cellcolor{underest} 0.65 &\cellcolor{underest}  0.50 &\cellcolor{underest}  0.38 \\
     0.0  &\cellcolor{match} 1.00 &\cellcolor{match}  1.00 &\cellcolor{match}  1.00 &\cellcolor{match}  0.97 &\cellcolor{underest}  0.87 &\cellcolor{underest}  0.72 &\cellcolor{underest} 0.57 &\cellcolor{underest}  0.43 \\
     3.0  &\cellcolor{match}  1.00 &\cellcolor{match}  1.00 &\cellcolor{overest}  1.07 &\cellcolor{overest}  1.13 &\cellcolor{match}  1.03 &\cellcolor{underest}  0.86 &\cellcolor{underest}  0.67 &\cellcolor{underest}  0.51 \\
     6.0  &\cellcolor{match}  1.00 &\cellcolor{overest}  1.07 &\cellcolor{overest}  1.48 &\cellcolor{overest}  1.56 &\cellcolor{overest}  1.38 &\cellcolor{overest}  1.12 &\cellcolor{underest}  0.86 &\cellcolor{underest}  0.64 \\
    10.0  &\cellcolor{overest}  1.09 &\cellcolor{overest}  2.66 &\cellcolor{overest}  3.07 &\cellcolor{overest}  2.77 &\cellcolor{overest}  2.25 &\cellcolor{overest}  1.74 &\cellcolor{overest}  1.31 &\cellcolor{match}  0.96 \\
    20.0  &\cellcolor{overest} 23.39 &\cellcolor{overest} 19.02 &\cellcolor{overest} 14.68 &\cellcolor{overest} 10.99 &\cellcolor{overest}  8.08 &\cellcolor{overest}  5.86 &\cellcolor{overest}  4.22 &\cellcolor{overest}  3.02 \\
        \bottomrule
\end{tabular}

\end{table}

This behaviour changes for larger arrays and is exemplified by the normalised local diversity in Figs. \ref{fig:loc_div_rayleigh} and \ref{fig:loc_div_rice} for a Rayleigh and Rician channel with $\mathcal{K}$-factor \SI{10}{\deci\bel}, respectively.
Tab. \ref{tab:nld_rice} summarises the results for a probability of \num{e-6} over different $\mathcal{K}$-factors and number of antennas $M$.
The normalisation is achieved by dividing the local diversity with the number of antennas.
Hence, once the normalised local diversity attains unity, the classic diversity of $M$ for large \gls{snr} is achieved.
Therefore, the normalised local diversity can be interpreted as relative error between a lower tail approximation and the actual steepness of the effective gain \gls{cdf} at the chosen probability.

\section{Discussion} \label{sec:discussion}

\subsection{Validity of Lower Tail Approximations}
The relative error of diversity is provided in Tab. \ref{tab:nld_rice}, revealing three different connected regions.
The first region (green) is covering small $\mathcal{K}$-factors for small systems, where the normalised local diversity is close to unity. 
A lower tail approximation will give reasonable results for \gls{ur}-relevant statistics. 

The second region (blue) belongs to Rayleigh fading and smaller $\mathcal{K}$-factors for an increasing number of antennas.
In this case, the local diversity has not yet converged to unity and a lower tail approximation will overestimate the performance accordingly.
E.g. a 64 antenna element array in Rayleigh fading at a probability of \SI{e-6} will only provide the performance predicted by the asymptotic regime of a 32 antenna system.
For large systems, only significant deterministic components will provide superelevation in the region of interest.

The last region (red) belongs to large $\mathcal{K}$-factors, where the local diversity is larger than the diversity, e.g. an environment with a $\mathcal{K}$-factor of \SI{10}{\deci\bel} and 4 antennas presents a local diversity of $4 \cdot 3.07 \approx 12$ in the superelevated probability region.
The superelevation moves towards smaller probabilities for an increasing number of antennas.
Overall, the deterministic component of a Rician fading environment plays a role for every $\mathcal{K}$-factor for large antenna systems and a growing $\mathcal{K}$-factor increases the local diversity.

Tab. \ref{tab:nld_rice} demonstrates clearly that a low tail approximation is giving misleading results for the effective channel gain of massive \gls{mimo} systems in Rayleigh and Rician fading.

\begin{table}[tb]
    \centering
    \caption{Analytic fading margins in \si{\deci\bel} at \num{e-6} probability.}
    \label{tab:fm_rice}
    \setlength{\tabcolsep}{4pt}
    \begin{tabular}{r
SSSSSSSS
    				}
        \toprule
        $\mathcal{K}$ & \multicolumn{7}{c}{Number of Antennas ($M$)} \\
        \cmidrule(lr){2-9}
        [\si{\deci\bel}] & {\centering 1} & {\centering 2} & {\centering 4} & {\centering 8} & {\centering 16} & {\centering 32} & {\centering 64} & {\centering 128} \\
        \midrule
    $-\infty$  & 58.4 & 30.7 & 17.1 & 10.2 &  6.5 &  4.3 &  2.9 &  2.0 \\
     0.0  & 57.6 & 29.7 & 16.0 &  9.3 &  5.7 &  3.7 &  2.5 &  1.7 \\
     3.0  & 55.3 & 27.3 & 13.9 &  7.8 &  4.9 &  3.2 &  2.1 &  1.5 \\
     6.0  & 49.2 & 21.3 & 10.2 &  6.0 &  3.8 &  2.5 &  1.7 &  1.2 \\
    10.0  & 27.2 & 10.4 &  5.9 &  3.7 &  2.5 &  1.7 &  1.2 &  0.8 \\
    20.0  &  3.5 &  2.3 &  1.6 &  1.1 &  0.8 &  0.5 &  0.4 &  0.3 \\
        \bottomrule
    \end{tabular}
\end{table}

\subsection{Array Deployment Strategies}
In the following the impact of some array deployment strategies for \gls{urllc} applications is discussed.
We relate the local diversity to the fading margin, another tangible figure of merit.
The fading margin is describing the gap between the median of the effective channel gain distribution and a target outage probability \cite{abraham_fading_2021}.
It has been evaluated for the same parameters as the normalised local diversity and the result is presented in Tab. \ref{tab:fm_rice}.
This complementary perspective highlights the return on investment of extra power or antenna gain, to improve the reliability of a system.

Regarding each column in the table shows, that every increase of the deterministic component will reduce the margin, thereby improving the robustness of the system.
Hence, it is worthwhile to compare a larger co-located system with a smaller $\mathcal{K}$-factor to smaller spatially distributed deployments.

It can be assumed that a distributed deployment will have at least one subarray closer to a user, giving a larger $\mathcal{K}$-factor.
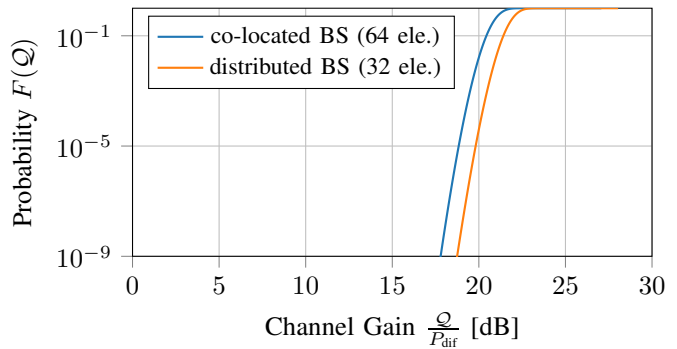
\begin{figure}[tb]
	\begin{tikzpicture}
\begin{axis}[
legend pos=north west,
tick align=outside,
tick pos=left,
xlabel={Channel Gain $\frac{\mathcal{Q}}{P_\text{dif}}$ [dB]},
xmajorgrids,
xmin=0, xmax=30,
ylabel={Probability $F(\mathcal{Q})$},
ymode=log,
ymajorgrids,
ymin=1e-9, ymax=1,
width=.95\columnwidth, height=.55\columnwidth,
]
\addplot [thick, color0] table[x expr=\thisrowno{0}+10*log10(128), y index=2] {data/K_0-M_64.csv};
\addlegendentry{\small{co-located BS (64 ele.)}};
\addplot [thick, color1] table[x expr=\thisrowno{0}+10*log10(160), y index=2] {data/K_6-M_32.csv};
\addlegendentry{\small{distributed BS (32 ele.)}};
\end{axis}
\end{tikzpicture}
	\caption{\glspl{cdf} of a co-located 64 antenna base station with $\mathcal{K}$-factor \SI{0}{\deci\bel} (blue) and the closest distributed 32 antenna base station (orange) with $\mathcal{K}$-factor \SI{6}{\deci\bel}. The stronger deterministic component of the channel in the distributed base station case compensates for the reduced number of antennas, resulting in a similar local diversity at \num{e-6}, giving a slight advantage with respect to the mean of the channel gain.}
	\label{fig:bs_example}
	\vspace{-8pt}
\end{figure}
As an example: a co-located uncorrelated 64 antenna base station in a Rician fading environment with $\mathcal{K} = \SI{0}{\deci\bel} = 1$ would require a fading margin of \SI{2.5}{\deci\bel} at an outage probability for \num{e-6}.
The mean of the effective channel gain is $64 * (1 + 1) P_\text{dif} = 128 P_\text{dif}$.
Instead, placing two non-cooperating uncorrelated 32 antenna \glspl{bs} into that environment, which reduces the length of the deterministic path to a half for a user, could increase the $\mathcal{K}$-factor by \SI{6}{\deci\bel}.
The closer base stations would then require a fading margin of \SI{2.5}{\deci\bel} at an outage probability for \num{e-6}.
For this setting the mean gain would be $32 * (4 + 1) P_\text{dif} = 160 P_\text{dif}$.

The \glspl{cdf} of both deployments are shown in Fig. \ref{fig:bs_example}, where both slopes of have not yet converged to the asymptotic behaviour of the lower tail in the \gls{ur}-relevant regime.
In this toy example, distributed base stations requiring the same amount of hardware would give equal fading margins and increase the mean effective gain compared to the co-located case.
Hence, not only capacity improvements can be achieved by densification of base stations, but
\gls{ur}-relevant statistics can improve too without increasing the amount of deployed hardware.

In a more general situation, for fading environments with deterministic propagation components, the number of antenna elements per base station influences where the normalised local diversity shows superelevation.
We notice further, in a pure Rayleigh fading environment, increasing the number of base station antennas gives diminishing returns (see Fig. \ref{fig:loc_div_rayleigh}).

\subsection{Inferring \gls{ur}-relevant Statistics?}
So, how can we infer the system behaviour of events that barely ever happen?
Given a limited number of measurable samples from each antenna element, how could the \gls{ur}-relevant statistics be analysed in real world systems?

There are two basic approaches for \gls{ur} antenna arrays:
\subsubsection{Element Statistics}
The first is based on collection of antenna element observations, estimation of each distribution and careful modeling of correlation properties.
Antenna elements that belong to the same local area could be lumped into a single distribution to make more samples available.
Post-processing of the resulting distributions with combination strategies like \gls{sc} or \gls{mrc} result in a \gls{cdf} to be evaluated in the \gls{ur}-relevant regime. 
In case of \gls{sc}, it is not necessary to have a reliable estimate of the antenna element \glspl{cdf} in that regime, but rather in the regime resulting from the $M$-th root of the target outage probability.
This follows from the maximum order statistic \cite{yang_order_2011} for the strongest constituent, being the $M$-th power of the element \gls{cdf}.
Since \gls{mrc} will give a better combined gain than \gls{sc}, using a \gls{sc} result allows to bound the system behaviour in the \gls{ur}-relevant regime based on reliable estimates of the element \glspl{cdf}.

\subsubsection{Combined Statistics}
The second approach implements a specific combination strategy, evaluating the \gls{ur}-relevant statistics directly.
This includes intrinsically antenna correlation, avoiding the necessity of explicit characterisation.
Unfortunately, this strategy requires prohibitively many observations.
Even for the first approach a lot of samples are necessary, but the antenna element observations do not need to be observed in the \gls{ur}-relevant regime directly, since this regime only matters for the effective channel gain!
Furthermore, the correlation is expected to depend to a lesser extent on the combined channel stationarity, allowing them to be studied in more detail with help of all antenna element observations.

\subsection{Correlated Channels}
Even though this manuscript demonstrated a local diversity based analysis for uncorrelated systems, the same ideas can be transferred to correlated antenna arrays.
Analytic results can be derived from the effective channel gain \gls{pdf} and \gls{cdf} of the correlated system, to avoid
Monte Carlo simulations that depend on a large amount of observations to provide reasonable insight.


\section{Conclusion}

Acquisition of \gls{ur}-relevant channel statistics is difficult to achieve in practical situations, because the number of required observations is tremendous. 
Ultimately, the spatial, spectral and temporal stationarity of the radio channel restricts the collection of a sufficient number of observations.
The approach of using the asymptotic lower tail behaviour, to avoid determination of a specific fading distribution, can be used for small arrays up to four antennas in low $\mathcal{K}$-factor Rician fading.
Systems that provide large diversity, require consideration of the local diversity in the \gls{ur}-relevant regime.
There, the asymptotic behaviour applies to probabilities beyond the \gls{ur}-relevant regime only.
Normalisation of the local diversity with the number of antenna elements in an array gives a relative deviation from the classic diversity.
Furthermore, the local diversity opens up for performance evaluation, where measurements of correlated antenna systems can be compared to an uncorrelated optimum.

For fast and numerically stable calculations, the distribution functions and local diversity of the effective gain of an uncorrelated antenna array in Rician fading can be formulated on the basis of the complementary Marcum-Q function.
Evaluation of the fading margin and distribution mean reinforces that a dense deployment of smaller base stations with the potential for increased deterministic radio channels is preferable over very large co-located systems, not only improving system capacity but also robustness.

\appendix
\subsection{\gls{cdf} of the Effective Power Gain} \label{sec:app-cdf}

The \emph{non-central gamma distribution} has \gls{pdf} $w_\rho(x; \alpha, \mu)$ for index $\rho$, scale $\alpha$, and non-centrality $\mu$ \cite[(1.47')]{ruben_non-central_1974} for $x \ge 0$:
\begin{equation}
	w_\rho(x; \alpha, \mu) = \frac{1}{\alpha} e^{-\frac{x}{\alpha} - \mu} \left( \frac{x}{\alpha \mu} \right)^{\frac{1}{2}(\rho - 1)} I_{\rho-1} \left( 2 \sqrt{\frac{\mu x}{\alpha}}\right). \label{eqn:nc-gamma-pdf}
\end{equation}
The corresponding \gls{cdf} $W_\rho(x; \alpha, \mu)$ can be directly related to the definition of the complementary Marcum Q-function in Eqn. \eqref{eqn:marcum_p} by substitution of $t = \frac{x'}{\alpha}$ in the integral relation between \gls{cdf} and \gls{pdf}:
\begin{equation}
	 W_\rho(x; \alpha, \mu) = \int_0^x w_\rho(x'; \alpha, \mu) dx' = P_\rho(\mu, \frac{x}{\alpha}). \label{eqn:nc-gamma-cdf}
\end{equation}

The gain \gls{pdf} $f(\mathcal{Q}; P_\text{dif}, \mathcal{K}, M=1)$ of a single antenna Rician channel is readily available by using the Rician envelope \gls{pdf} from \cite[(5.3.7)]{durgin_space-time_2003}
applying the transformation to the power \gls{pdf} \cite[(5.2.1)]{durgin_space-time_2003} and replacing the power term of the deterministic component \cite[(5.3.8)]{durgin_space-time_2003}, resulting in:
\begin{align}
	f(\mathcal{Q};& P_\text{dif}, \mathcal{K}, M=1) \notag \\
	&= \frac{1}{P_\text{dif}} \exp \left(- \frac{\mathcal{Q}}{P_\text{dif}} - \mathcal{K} \right) I_0\left(2 \sqrt{\mathcal{K} \frac{\mathcal{Q}}{P_\text{dif}}} \right) \notag \\
	&= w_1(\mathcal{Q}; P_\text{dif}, \mathcal{K}). \label{eqn:app-single-antenna-pdf}
\end{align}
This \gls{pdf} is a special case of the non-central gamma distribution \gls{pdf} in Eqn. \eqref{eqn:nc-gamma-pdf} for index one, scale $P_\text{dif}$ with non-centrality $\mathcal{K}$.

For $M$ independent single antenna Rician channels with potentially differing $\mathcal{K}$-factors $\mathcal{K}_m \forall m \in [1, \cdots, M]$ the additive property of non-central gamma distributions can be used to get the \gls{pdf} of the effective channel gain.
The addition property allows to represent the sum of independent random variables with the same scale, potentially varying index and non-centrality as non-central gamma distribution \cite[(1.51)]{ruben_non-central_1974}.
The generalisation of Eqn. \eqref{eqn:app-single-antenna-pdf} for an $M$ antenna array follows a non-central gamma distribution of index $M$ and non-centrality $\sum_{m=1}^M \mathcal{K}_m$:
\begin{equation}
	 f(\mathcal{Q}; P_\text{dif}, \mathcal{K}, M) = w_{M} \left(\mathcal{Q}; P_\text{dif}, \sum_{m=1}^M \mathcal{K}_m \right) \label{eqn:app-multi-antenna-pdf}.
\end{equation}

Using Eqn. \eqref{eqn:nc-gamma-cdf} gives the \gls{cdf} of the effective channel gain based on the inverse Marcum Q-function:
\begin{equation}
	F(\mathcal{Q}; P_\text{dif}, \mathcal{K}, M) = P_M \left( \sum_{m=1}^M \mathcal{K}_m, \frac{\mathcal{Q}}{P_\text{dif}} \right).
\end{equation}

\IEEEtriggeratref{9} 
\printbibliography

\end{document}